\documentclass[10pt,letterpaper]{article}
\usepackage{cmap}
\usepackage[T1]{fontenc}
\usepackage{spconf,amsmath,amssymb,graphicx,booktabs,array,tabularx,cite,url}
\usepackage[table]{xcolor}
\usepackage[protrusion=false,expansion=true]{microtype}
\definecolor{rowgray}{gray}{0.93}
\definecolor{revegray}{gray}{0.93}
\newcommand{\revetablestyle}{%
  \fontsize{9}{11.3}\selectfont
  \setlength{\tabcolsep}{2pt}%
  \setlength{\heavyrulewidth}{0.6pt}%
  \setlength{\lightrulewidth}{0.35pt}%
  \setlength{\aboverulesep}{1.2pt}%
  \setlength{\belowrulesep}{1.2pt}%
  \renewcommand{\arraystretch}{1.00}%
}
\newcommand{\thcell}[1]{\multicolumn{1}{c}{\bfseries\shortstack{#1}}}
\makeatletter
\def\@maketitle{\newpage
 \null\vskip 2em
 \begin{center}
 {\fontsize{14}{16}\selectfont\bfseries\@title\par}
 \vskip 1.5em
 {\fontsize{12}{14}\selectfont\lineskip .5em
\begin{tabular}[t]{c}\@name\\\@address\end{tabular}\par}
\end{center}\par\vskip 1.5em
}
\makeatother
\title{REVE: EFFICIENT HALLUCINATION CORRECTION FOR\\LARGE AUDIO-LANGUAGE MODELS VIA REUSED ENCODER STATES}
\name{Hongjin Song, Jiasheng Kuang, Xinyu Yang, Qiuyu Fang, Ziyu Wu, Guowu Tan, Xiang Xie$^{*}$}
\address{Beijing Institute of Technology, Zhuhai, Guangdong, China\\
{\fontsize{9}{11}\selectfont $^{*}$Corresponding author: Xiang Xie}}
\begin{document}
\ninept
\normalsize
\maketitle
\begin{abstract}
Large audio-language models may mention acoustic events that are absent from the input. A separate audio event detector can verify these mentions, but doing so requires a second audio encoder and a separate forward pass. We propose Reused Encoder States for Verifying Events (REVE), a lightweight method that uses states already computed by the target model. One readout summarizes class scores across audio frames, while another uses pooled states from four consecutive frame intervals. Class-aware score fusion combines their outputs to verify generated event mentions without encoding the audio again. On AudioSet, REVE removes 92.9\% of label-unsupported mentions under a faithful-mention recall constraint. With fewer added parameters and no second audio-encoding pass, REVE achieves a reduction comparable to those of CED-Tiny and CED-Base. Its complete verification latency is about $1/18$ of the CED-Base path. Results on controlled DESED mixtures and different target-model architectures further confirm the effectiveness of encoder-state reuse.
\end{abstract}

\section{Introduction}

Large audio-language models (LALMs) usually connect an audio encoder and a large language model through a modality projector. This design supports tasks such as audio question answering and open-ended audio captioning. For example, Qwen2-Audio~\cite{qwen2audio} first uses a Whisper-based audio encoder~\cite{whisper} to extract acoustic representations. A modality projector then maps these representations into audio embeddings that are compatible with the input space of the language model. The language model combines the audio embeddings with a text instruction to generate an answer or caption. Despite their strong audio understanding, LALMs may still describe acoustic events that are absent from the input. For example, a model may mention thunder in a rain-only clip or music in a speech recording. Such descriptions are often plausible. It is therefore difficult to verify them using only the generated text or token probabilities. Prior work reduces audio hallucinations through audio-aware decoding~\cite{aad2025}, activation steering~\cite{avs2025}, preference alignment~\cite{aha2026}, and broader evaluation frameworks~\cite{halluaudio2026}. However, caption hallucinations often occur at the event-mention level. A caption-level score cannot locate the exact unsupported text span. Changes to model training or decoding may also be difficult to apply to a deployed model.

A direct post-hoc solution first extracts event mentions from the generated caption, maps them to classes in an audio event ontology, and then queries a separate detector such as CED~\cite{ced}. This process provides class-level acoustic evidence and links each detection result to a specific caption span. It can therefore retain supported event mentions and remove unsupported ones. However, CED-Base introduces a high parameter cost and requires a second audio-encoder forward pass. CED-Tiny substantially reduces the detector size, but it still extracts acoustic features and encodes the waveform again. This additional path increases verification latency. Since the target model has already encoded the audio before generating the caption, we ask whether its existing encoder representations can provide event-presence evidence without running another audio model.

To this end, we propose \textbf{Reused Encoder States for Verifying Events (REVE)}. It uses two readouts: one summarizes frame-level class scores from the pre-projector states, and the other reads out the means of four contiguous, equal-length frame intervals. A class-aware calibrator fuses the two scores and produces the final event-presence score for each ontology class. REVE maps event spans in the generated caption to ontology classes and uses these scores to verify the corresponding events.

Our contributions are threefold. First, we formulate post-hoc audio caption correction as ontology-aligned event verification and introduce a two-scale verifier that reuses encoder states. Second, we evaluate hallucination reduction under a faithful-mention recall constraint and report the added computation and storage costs of the verification module. Third, source-disjoint controlled DESED experiments and cross-model experiments show that REVE generalizes across acoustic domains and LALM architectures.

\section{Method}

Fig. 1 shows the full caption generation and event verification process. REVE reuses frame-level encoder states before the modality projector as acoustic evidence. It maps event spans in the generated caption to an audio event ontology and retains only mentions supported by the corresponding class scores.

\begin{figure*}[t]
\centering
\includegraphics[width=\textwidth]{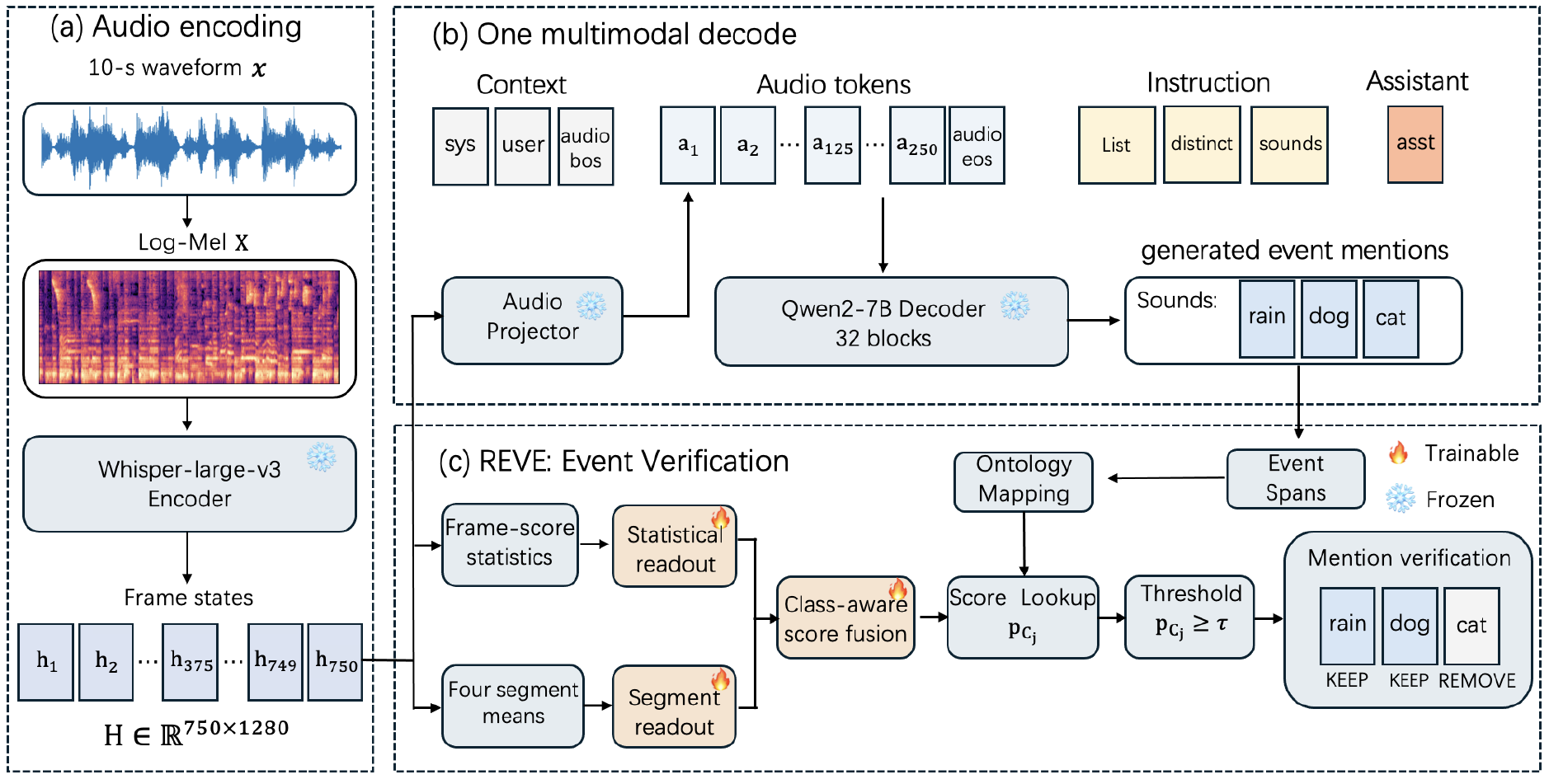}
\caption{Caption generation and event verification. (a) The audio encoder extracts frame-level states $H$. (b) The target LALM performs one standard multimodal decoding pass. (c) REVE combines frame-level score statistics with four temporal segment representations, fuses the two signals in a class-aware manner, and removes unsupported event mentions.}
\label{fig:1}
\end{figure*}

\subsection{Encoder State Reuse}

Let $x$ denote the input waveform and $q$ the captioning instruction. Let $E$, $P$, and $G$ denote the audio encoder, modality projector, and language-model decoder of the target LALM, respectively. The audio encoder first extracts frame-level states from a log-Mel spectrogram:

\begin{equation}
H=E\bigl(\operatorname{LogMel}(x)\bigr)
=[h_1,\ldots,h_T]^\top\in\mathbb{R}^{T\times d}.
\end{equation}

where $T$ is the number of encoder frames, $d$ is the dimension of each encoder state, and $h_t\in\mathbb{R}^{d}$ is the state at frame $t$. The modality projector then converts $H$ into audio tokens that are compatible with the language model:

\begin{equation}
Z=P(H),\qquad Z\in\mathbb{R}^{L\times d_G}.
\end{equation}

where $L$ is the number of audio tokens and $d_G$ is the token dimension. The language model generates caption $C$ from $Z$ and instruction $q$:

\begin{equation}
C=G(Z,q).
\end{equation}

As shown in Fig. 1(a)--(b), the original caption generation path and REVE share the same encoder states $H$. REVE can therefore extract acoustic evidence directly from the existing states without reloading or re-encoding the input audio.

\subsection{Ontology Mapping}

The language model uses free-form event descriptions, whereas the readouts use fixed classes. REVE derives matching forms from the official AudioSet class names by normalizing case and separators and removing parenthetical qualifiers. Whole-word matching identifies event spans in caption $C$:

\begin{equation}
\mathcal{M}(C)=\{(s_j,c_j)\}_{j=1}^{J},
\end{equation}

where $s_j$ is an event span and $c_j\in\{1,\ldots,K\}$ is its ontology class. A matched node maps directly to a readout class or to its nearest represented ancestor. Unmapped spans are not edited. Reference labels use the same mapping.

\subsection{Two-Scale Event Scoring}

Mean pooling gives a compact clip-level representation but may weaken evidence for short events. To capture both global score distributions and local temporal information, REVE combines frame-level score statistics with four contiguous, equal-length temporal segments. We first mean-pool all encoder frames to obtain the base clip representation:

\begin{equation}
e=\frac{1}{T}\sum_{t=1}^{T}h_t,
\qquad e\in\mathbb{R}^{d}.
\end{equation}

A multi-label readout defines the base class logit $g_c=w_c^\top e+b_c$ and the frame-level logit $a_{t,c}=w_c^\top h_t+b_c$. For each mentioned class, we summarize the temporal distribution of its frame-level scores as

\begin{equation}
q_c=[g_c,\operatorname{std}_t a_{t,c},\max_t a_{t,c}],
\end{equation}

and obtain $s_c^{\rm stat}=\sigma(\alpha_c^\top q_c+\beta_c)$. In parallel, we divide the $T$ encoder frames into four contiguous temporal segments $\mathcal I_l$ and retain the mean representation of each segment:

\begin{gather}
e_l=|\mathcal I_l|^{-1}\sum_{t\in\mathcal I_l}h_t,\quad l=1,\ldots,4,\\
s^{\rm seg}=\sigma\!\left(W_4[e_1;\ldots;e_4]+b_4\right).
\end{gather}

Both readouts are trained on AudioSet-balanced-train with class-balanced binary cross-entropy. Let $\ell(\cdot)$ denote the logit function and define $z_c=[\ell(s_c^{\rm stat}),\ell(s_c^{\rm seg})]^\top$. The final event-presence score is

\begin{equation}
p_c=\sigma\!\left(\theta^\top z_c+\delta_c+
\gamma_c\frac{\mathbf 1^\top z_c}{2}+b\right).
\end{equation}

Here, $\theta$ and $b$ are shared across classes, while $\delta_c$ and $\gamma_c$ are the class-specific offset and slope. This regularized class-aware calibrator adds only 1,057 coefficients. All parameters of the target LALM remain frozen during training. The original readout that uses only one mean-pooled representation serves as the \emph{Mean Readout} ablation.

\subsection{Event Mention Correction}

After ontology mapping, REVE scores only the classes mentioned in the caption. It produces score $p_{c_j}$ for each event span $s_j$. This candidate-sparse implementation is equivalent to scoring all $K$ classes. The threshold decision is defined as

\begin{equation}
r_j(\tau)=\mathbb{I}\left[p_{c_j}\geq\tau\right].
\end{equation}

where $\mathbb{I}[\cdot]$ is the indicator function and $\tau$ is the event-presence threshold. The event span is retained when $r_j=1$ and removed when $r_j=0$. The corrected caption is

\begin{equation}
C'=\operatorname{Edit}
\left(C,\left\{s_j\mid r_j(\tau)=0\right\}\right).
\end{equation}

REVE edits only event spans with an ontology class score. All other text remains unchanged.

\section{Experimental Setup}

\subsection{Datasets and Models}

\textbf{AudioSet.} We train the two readouts on AudioSet-balanced-train~\cite{audioset}. For evaluation, we divide 1,000 AudioSet-eval clips into non-overlapping development and test sets of 500 clips each. Only the development set is used to select the class-aware calibrator and threshold. All methods use the same captions, event spans, ontology mappings, and reference labels.

\textbf{DESED mixtures.} We create 1,000 ten-second mixtures from isolated DESED foreground events~\cite{desed2021}. They cover eight event families and contain one to three distinct events without background noise. Development and test sets contain 500 mixtures each and use disjoint source waveforms. The mixing records provide exact event-presence labels.

\textbf{Models.} Our main target is Qwen2-Audio-7B-Instruct~\cite{qwen2audio}. For each ten-second clip, its encoder produces $H\in\mathbb{R}^{750\times1280}$ and the projector produces 250 audio tokens. Captions use greedy decoding. We also test Qwen2.5-Omni-7B and SALMONN-13B~\cite{qwen25omni,salmonn}.

\subsection{Baselines}

We compare signals from generation confidence, internal target-model states, and external audio models. \textbf{Token Confidence} uses probabilities aligned with each event mention. \textbf{Decoder Probe} uses the last-layer language-model states. \textbf{Mean Readout} uses one mean-pooled encoder representation. External baselines are \textbf{CLAP}~\cite{clap}, \textbf{AST}~\cite{ast}, \textbf{PANN Cnn14}~\cite{panns}, and the \textbf{CED-Tiny} and \textbf{CED-Base} audio taggers~\cite{ced}. These models reprocess the audio, whereas REVE reuses target-model states.

\subsection{Evaluation Protocol}

Let $\mathcal{I}$ be the set of evaluated event mentions in the original captions. Let $y_i\in\{0,1\}$ indicate whether mention $i$ is supported by the reference labels. For scoring method $m$ and threshold $\tau$, let $r_i^{(m)}(\tau)\in\{0,1\}$ indicate whether the mention is retained after pruning. We define residual false-mention density and faithful-mention recall as

\begin{gather}
D_m(\tau)=\frac{\sum_{i\in\mathcal I}(1-y_i)r_i^{(m)}(\tau)}{|\mathcal I|},\\
R_m(\tau)=\frac{\sum_{i\in\mathcal I}y_i r_i^{(m)}(\tau)}{\sum_{i\in\mathcal I}y_i}.
\end{gather}

$D_m$ is the fraction of all original evaluated mentions that are false but remain after pruning. Lower values are better. $R_m$ is the fraction of reference-supported mentions that remain after pruning.

We report relative hallucination reduction as

\begin{equation}
Q_m(\tau)=1-\frac{D_m(\tau)}{D_{\mathrm{no\ pruning}}}.
\end{equation}

The reduction directly measures the fraction of original false mentions that are removed. However, it must be reported together with faithful-mention recall, because deleting every mention would otherwise give a 100\% reduction. We therefore maximize the reduction under the development-set recall constraint below.

For each score-based method, we first obtain the set of development thresholds that satisfy the recall constraint:

\begin{gather}
\mathcal F_m=\{\tau\in\mathcal T_m:R_m^{\mathrm{dev}}(\tau)\geq0.75\},\\
\tau_m^*=\max\mathcal F_m.
\end{gather}

Class-balanced training makes the acoustic outputs decision scores rather than calibrated probabilities. We therefore select $\tau$ on the development set instead of fixing it at $0.5$. REVE calibration and threshold selection use only grouped out-of-fold predictions from the development set. The complexity analysis includes only the added verification cost for a ten-second clip and excludes the shared target LALM. The data-processing, evaluation, and complexity-profiling scripts will be released on GitHub.

\begin{figure*}[t]
\centering
\includegraphics[width=\textwidth]{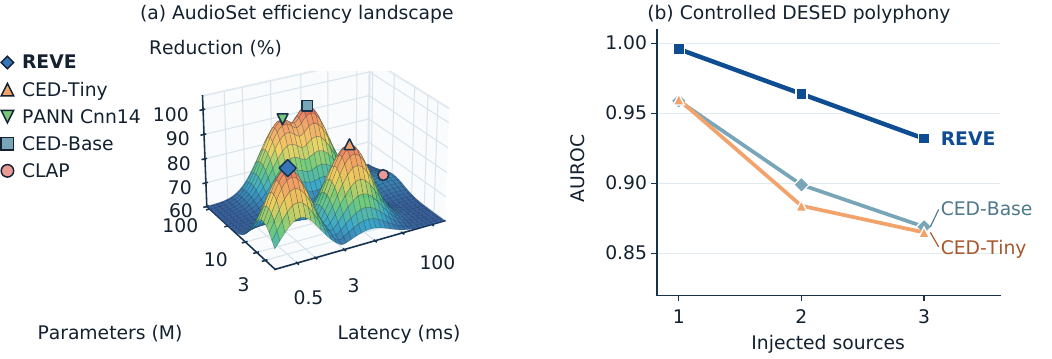}
\caption{Efficiency and controlled-mixture results. (a) AudioSet reduction versus complete added latency and parameters (log scales). Markers are measurements; the smoothed surface is a visual guide. (b) AUROC across controlled DESED polyphony levels.}
\label{fig:2}
\end{figure*}

\section{Results}

\subsection{AudioSet Results}

Table 1 reports results under the development-set recall constraint. REVE lowers residual density from $0.844$ to $0.060$ and removes $92.9\%$ of false mentions at $0.756$ recall. This is comparable to the $92.7\%$ and $93.8\%$ reductions of CED-Tiny and CED-Base, without a separate audio-encoder pass.

\par\smallskip
\noindent\begin{minipage}{\columnwidth}
\noindent\refstepcounter{table}\textbf{Table \thetable.} AudioSet results at a development-set recall target of 0.75. REVE is shown in bold; underlining marks a best result only when it is achieved by another method.
\par\vspace{3pt}
\begingroup\revetablestyle
\begin{tabularx}{\columnwidth}{@{}>{\columncolor{white}[0pt][\tabcolsep]\raggedright\arraybackslash}Xcc>{\columncolor{white}[\tabcolsep][0pt]}c@{}}
\toprule
\textbf{Method} & \thcell{Recall $\uparrow$} & \thcell{Residual\\density $\downarrow$} & \thcell{Reduction $\uparrow$}\\
\midrule
No Correction & 1.000 & 0.844 & 0.0\%\\
Token Confidence & \underline{0.810} & 0.563 & 33.3\%\\
Decoder Probe & 0.747 & 0.207 & 75.5\%\\
Mean Readout & 0.774 & 0.103 & 87.8\%\\
CLAP & 0.785 & 0.301 & 64.3\%\\
PANN Cnn14 & 0.791 & 0.077 & 90.9\%\\
AST & 0.802 & 0.068 & 91.9\%\\
CED-Tiny & 0.786 & 0.061 & 92.7\%\\
CED-Base & 0.758 & \underline{0.052} & \underline{93.8\%}\\
\rowcolor{rowgray}
\textbf{REVE} & \textbf{0.756} & \textbf{0.060} & \textbf{92.9\%}\\
\bottomrule
\end{tabularx}
\endgroup
\end{minipage}
\par\vspace{0.5\baselineskip}

Table 2 compares the added deployment cost with CED. CED-Tiny reduces CED-Base from $85.71$M to $5.50$M parameters and from $10.32$ to $8.27$ ms complete latency. Both still extract features and encode the waveform again. REVE adds $3.38$M parameters. Its GPU computation takes $0.29$ ms and its synchronized complete latency is $0.57$ ms, about $1/18$ of the CED-Base path.

\par\smallskip
\noindent\begin{minipage}{\columnwidth}
\noindent\refstepcounter{table}\textbf{Table \thetable.} Incremental cost for a 10-s clip on an NVIDIA L20Y in FP32. GPU forward uses CUDA-event timing; CED complete latency also includes feature extraction and transfer. REVE is shown in bold.
\par\vspace{3pt}
\begingroup\revetablestyle
\begin{tabularx}{\columnwidth}{@{}>{\columncolor{white}[0pt][\tabcolsep]\raggedright\arraybackslash}Xcc>{\columncolor{white}[\tabcolsep][0pt]}c@{}}
\toprule
\textbf{Method} & \thcell{Parameters} & \thcell{GPU forward\\(ms)} & \thcell{Complete path\\(ms)}\\
\midrule
CED-Tiny & 5.50M & 4.20 & 8.27\\
CED-Base & 85.71M & 5.46 & 10.32\\
\rowcolor{rowgray}
\textbf{REVE} & \textbf{3.38M} & \textbf{0.29} & \textbf{0.57}\\
\bottomrule
\end{tabularx}
\endgroup
\end{minipage}
\par\vspace{0.5\baselineskip}

Fig. 2(a) shows that REVE reaches a $92.9\%$ reduction with only $0.57$ ms of added latency, while PANN, CED, and CLAP are slower.

\subsection{Ablation Study}

Table 3 shows that the two readouts provide complementary evidence. Shared fusion raises mention AUROC from about $0.91$ to $0.931$ and lowers residual density to $0.082$. Class-aware calibration keeps AUROC nearly unchanged but lowers residual density to $0.060$, which supports more accurate decisions with one threshold across classes.

\par\smallskip
\noindent\begin{minipage}{\columnwidth}
\noindent\refstepcounter{table}\textbf{Table \thetable.} Component ablation on AudioSet. Residual density is measured at the development-selected recall target of 0.75. REVE is shown in bold; underlining marks a best result only when it is achieved by another variant.
\par\vspace{3pt}
\begingroup\revetablestyle
\begin{tabularx}{\columnwidth}{@{}>{\columncolor{white}[0pt][\tabcolsep]\raggedright\arraybackslash}Xc>{\columncolor{white}[\tabcolsep][0pt]}c@{}}
\toprule
\textbf{Variant} & \thcell{Mention\\AUROC $\uparrow$} & \thcell{Residual\\density $\downarrow$}\\
\midrule
Mean Readout & 0.905 & 0.103\\
Statistics only & 0.907 & 0.095\\
Segments only & 0.909 & 0.095\\
Shared fusion & \underline{0.931} & 0.082\\
\rowcolor{rowgray}
\textbf{REVE} & \textbf{0.930} & \textbf{0.060}\\
\bottomrule
\end{tabularx}
\endgroup
\end{minipage}
\par\vspace{0.5\baselineskip}

\subsection{Controlled DESED Results}

We use controlled DESED mixtures to avoid the missing-label ambiguity of AudioSet. The mixing records give exact presence labels for all eight event families.

Table 4 reports $0.959$ AUROC and $0.915$ AP for REVE. The strongest external detector, CED-Base, reaches $0.901/0.791$.

\par\smallskip
\noindent\begin{minipage}{\columnwidth}
\noindent\refstepcounter{table}\textbf{Table \thetable.} Source-presence diagnostic on controlled DESED mixtures. REVE is shown in bold.
\par\vspace{3pt}
\begingroup\revetablestyle
\begin{tabularx}{\columnwidth}{@{}>{\columncolor{white}[0pt][\tabcolsep]\raggedright\arraybackslash}Xc>{\columncolor{white}[\tabcolsep][0pt]}c@{}}
\toprule
\textbf{Verification method} & \thcell{Micro-AUROC $\uparrow$} & \thcell{Micro-AP $\uparrow$}\\
\midrule
CLAP & 0.821 & 0.644\\
PANN Cnn14 & 0.847 & 0.707\\
AST & 0.887 & 0.761\\
CED-Tiny & 0.896 & 0.779\\
CED-Base & 0.901 & 0.791\\
\rowcolor{rowgray}
\textbf{REVE (DESED)} & \textbf{0.959} & \textbf{0.915}\\
\bottomrule
\end{tabularx}
\endgroup
\end{minipage}
\par\vspace{0.5\baselineskip}

In Fig. 2(b), REVE obtains AUROCs of $0.996$, $0.964$, and $0.932$ for one to three sources. It remains above both CED variants at every source count.

\subsection{Cross-Model Results}

We also evaluate REVE on Qwen2.5-Omni-7B and SALMONN-13B~\cite{qwen25omni,salmonn} to test different target-model architectures.

\par\smallskip
\noindent\begin{minipage}{\columnwidth}
\noindent\refstepcounter{table}\textbf{Table \thetable.} Complete REVE across target LALMs at the development recall target of 0.75.
\par\vspace{3pt}
\begingroup\revetablestyle
\begin{tabularx}{\columnwidth}{@{}>{\raggedright\arraybackslash}Xccc@{}}
\toprule
\textbf{Target LALM} & \thcell{Mention\\AUROC $\uparrow$} & \thcell{Density\\before $\rightarrow$ after} & \thcell{Reduction}\\
\midrule
Qwen2-Audio-7B & 0.930 & 0.844 $\rightarrow$ 0.060 & 92.9\%\\
Qwen2.5-Omni-7B & 0.865 & 0.797 $\rightarrow$ 0.158 & 80.1\%\\
SALMONN-13B & 0.841 & 0.804 $\rightarrow$ 0.207 & 74.3\%\\
\bottomrule
\end{tabularx}
\endgroup
\end{minipage}
\par\vspace{0.5\baselineskip}

REVE reduces residual density to $0.158$ on Qwen2.5-Omni and $0.207$ on SALMONN, or by $80.1\%$ and $74.3\%$, respectively.

\section{Conclusion}

REVE corrects LALM captions using encoder states already available after caption generation. Its statistical and segment readouts remove $92.9\%$ of false mentions under the recall constraint, matching CED-Tiny and CED-Base. REVE uses fewer parameters, avoids another audio encoding pass, and adds about $1/18$ of CED-Base's complete verification latency. DESED and cross-model results confirm the effectiveness of encoder-state reuse.

\end{document}